# Effect of aging on the reinforcement efficiency of carbon nanotubes in epoxy matrix

A. Allaoui, P. Evesque and J. B. Bai[*]

*LMSSMAT CNRS UMR8579, Ecole Centrale Paris, 92290 Châtenay-Malabry, France*

**ABSTRACT**

The reinforcement efficiency of carbon nanotubes (CNTs) in epoxy matrix was investigated in the elastic regime. Cyclic uniaxial tensile tests were performed at constant strain amplitude and increasing maximum strain. Post-curing of the epoxy and its composite at a temperature close to the glass transition temperature allowed us to explore the effect of aging on the reinforcement efficiency of CNT. It is found that the reinforcement efficiency is compatible with a mean field mixture rule of stress reinforcement by random inclusions. It also diminishes when the maximum strain increased and this effect is amplified by aging. The decrease of elastic modulus with increasing cyclic maximum strain is quite similar to the one observed for filled elastomers with increasing strain amplitude, a phenomenon often referred as the Payne effect.

---

[*] Corresponding author. Phone: +33 41 13 13 16. E-mail address: jinbo.bai@ecp.fr (JinBo Bai).

The reinforcement of polymer matrices with carbon nanotubes (CNTs) has been the subject of a prolific literature, especially with epoxy as the matrix. Depending on the preparation method, the reinforcement potential of CNTs (theoretical Young modulus of 1 TPa) was partially [1,2] or fully exploited [3]. A review [4] has been published on the subject. However, to our knowledge, the effect of aging has never been explored. In this paper, the elastic properties of CNT reinforced epoxy composites and the effect of aging on the reinforcement efficiency of CNT were investigated by means of cyclic uniaxial tensile tests at increasing maximum strain and constant strain amplitude.

Multiwall carbon nanotubes (MWNT) with diameters in the range of 10-150 nm and average length of 10 μm, were produced by the CVD method [5]. A low viscosity diglycidyl ether of bisphenol-A (DGEBA) epoxy resin ($\eta \sim 0.7$ Pa.s) with triethylenetetramine (TETA) hardener (Epofix, Struers) was used. The stochiometric ratio was 100:12 (w/w) epoxy resin:hardener. Carbon nanotubes were dispersed in ethanol using magnetical agitation for 3 days. CNTs were filtered with a vacuum pump and then heated until complete removal of the solvent. The CNT powder was dispersed in epoxy resin using mechanical mixing for half an hour. The hardener was added and the mixture was mixed for another 5 min, vacuum pumped and cast into silicon moulds. The samples were cured at room temperature for 3 days followed by post-cure at 80 °C for 24 h. The dogbone samples (width ~6 mm, length ~30 mm, thickness ~2–3mm) were polished in order to eliminate the surface flaws. Six unfilled epoxy resin samples and seven composites epoxy + 0.5 wt.% MWNT were prepared.

Cyclic uniaxial tensile tests were performed at constant strain amplitude with an electromechanical testing stage (Instron 4505) equipped with a 1000 N force cell and an extensometer with 12.5 ± 2.5 mm gage length. A 10N pre-stress was applied to all samples. The strain amplitude was fixed to 0.1 % and the maximum strain was increased from 0.2 to 0.5 % to stay in the ''elastic'' regime. The crosshead speed was set to 0.5 mm/min. The elastic modulus was taken as the average slope of the stress–strain curve over three cycles at each maximum strain (0.2, 0.3, 0.4, and 0.5%). The samples were tested 3 days after their fabrication and tested again 1 month later to evaluate the effect of aging.

The elastic modulus of the unfilled epoxy and the composites tested after 3 days and after 1 month are plotted, respectively, in Fig. 1a and b. Physical aging allowed local rearrangement of the polymer chains leading to an increase of the modulus of the unfilled epoxy after 1 month. The decrease of the elastic modulus with increasing maximum cyclic strain is clearly much more marked in the case of the composite compared to the unfilled resin. This phenomenon is very similar to the substantial decrease of the modulus of carbon black filled rubbers with the increasing cyclic strain amplitude, which is often referred as the Payne effect [6]. We could expect the modulus to reach a plateau at small enough cyclic strain. However, this plateau is not observed yet in Fig. 1. In our case, the polymer was in the glassy state and the increase of the maximum strain at constant strain amplitude led to the decrease of the elastic modulus. It is worth noting that this softening was enhanced by aging as the modulus drop is higher after 1 month. Kraus [7] proposed a phenomenological model to describe the Payne effect as breakage and recovery of some weak physical bonds during strain oscillations. The model led to the following expression of the strain dependent elastic modulus

$$\frac{E(\varepsilon, age, CNT\ wt.\%) - E_\infty(age, CNT\ wt.\%)}{E_0(age, CNT\ wt.\%) - E_\infty(age, CNT\ wt.\%)} = \frac{1}{1 + (\varepsilon/\varepsilon_c(age, CNT\ wt.\%))^{2m}}$$

with $E_\infty(age, CNT\ wt.\%)$ and $E_0(age, CNT\ wt.\%)$ the elastic modulus, respectively, at infinite and zero strain, $\varepsilon_c(age, CNT\ wt.\%)$ is a characteristic strain at which the reinforcement drops to half of its initial value and m an adjustable parameter related to the filler aggregate structure and connectivity. In the original model, ε is the strain amplitude. In our case, the strain amplitude is constant, we thus considered ε as the maximum cyclic strain and fitted our data. A good agreement was found with m equal to 1. The set of fitted parameters for the composites tested after 3 days and after 1 month are summarized in Table 1. The effects of physical aging were manifested by the increase of $E_\infty$, corresponding to the increase of the unfilled epoxy modulus associated to rearrangement of the polymer chains, and the shift of the characteristic strain to lower values indicating an amplification of the modulus drop. To evaluate the CNT reinforcement efficiency, the rule of mixture with an orientation factor (with a value of 1/5 [8]) to take into account the random orientation of the CNT in the matrix was used. The CNT reinforcement efficiency was thus calculated, under a mean-field homogenization assumption with random distribution of fiber orientation, as

$$\frac{E_{composite} - (1-p)E_{epoxy}}{p/5}$$ with $E_{composite}$ and $E_{epoxy}$ are the elastic modulus of,

respectively, the composite and the unfilled epoxy, and p is the volume fraction of CNT. In the case of an effective reinforcement of CNT, the reinforcement efficiency should give an estimation of the average CNT Young modulus. The CNT reinforcement efficiency is plotted as a function of the maximum strain (Fig. 2). The reinforcement efficiency decreased with the maximum strain and this decrease was amplified by aging. The reinforcement efficiency at the lowest maximum strain applied gave a valuable evaluation of the average Young modulus of CNT around 580 GPa. A lower value than the theoretical 1 TPa is expected for CVD synthesized nanotubes due the presence of defects on this type of tubes. Cooper et al. [9] used the rule of mixture with an orientation factor to estimate the Young modulus of CNT from measured Raman band shift on strained CNT/epoxy composite films. They evaluated the modulus of SWNT to be around 1 TPa while they found a lower estimate for MWNT in the order of 0.3 TPa. The fitting parameters of the Kraus model provide the lower ($E_\infty$) and the upper bond ($E_0$) of the modulus. $E_0$ should correspond to the modulus of the composite with the maximum reinforcement achievable and $E_\infty$ to the modulus of the un-reinforced epoxy. The epoxy modulus was found to slightly decrease so that the zero-strain modulus was obtained using a linear fit. The zero-strain reinforcement efficiency can be evaluated using $E_0$ as the composite modulus and the extrapolated zero-strain epoxy modulus. It is worth noting that the reinforcement efficiency obtained is lower than the theoretical 1 TPa before aging and higher than this value after aging. This suggests that they could be an additional reinforcement effect after aging or a larger plateau modulus at sufficiently low cyclic strain. The explanation of this observation is still unclear. Further investigations may help to point out possible additional reinforcement due to CNT entanglement effect or polymer chains reduced mobility at the vicinity of CNT.

In this study, the reinforcement efficiency of CNTs fillers in epoxy matrix was assessed along with the influence of aging through cyclic tensile tests at constant strain amplitude and increasing maximum strain. The modulus of the composites was found to decrease with increasing maximum cyclic strain in a similar manner to the Payne effect in filled rubbers; that is the modulus decrease with increasing strain amplitude. The exponent of the Kraus model was found to be *m*=1. At sufficiently low strain, the reinforcement was effective and the average tensile modulus of these CVD CNTs can be deduced using the mixture law to be around 580 GPa. At higher strains, the CNT reinforcement efficiency diminished and this tendency was amplified by aging.

## LIST OF CAPTIONS

Figure 1 – Elastic modulus of unfilled epoxy and composites tested after (a) 3 days and (b) 1 month as a function of maximum cyclic strain. Solid lines are fits to Kraus model.

Figure 2 – CNT reinforcement efficiency as a function of maximum cyclic strain before and after aging.

Table 1 – Values of the parameters of the Kraus model fitted to the experimental tests performed on the composites after 3 days and after 1 month.

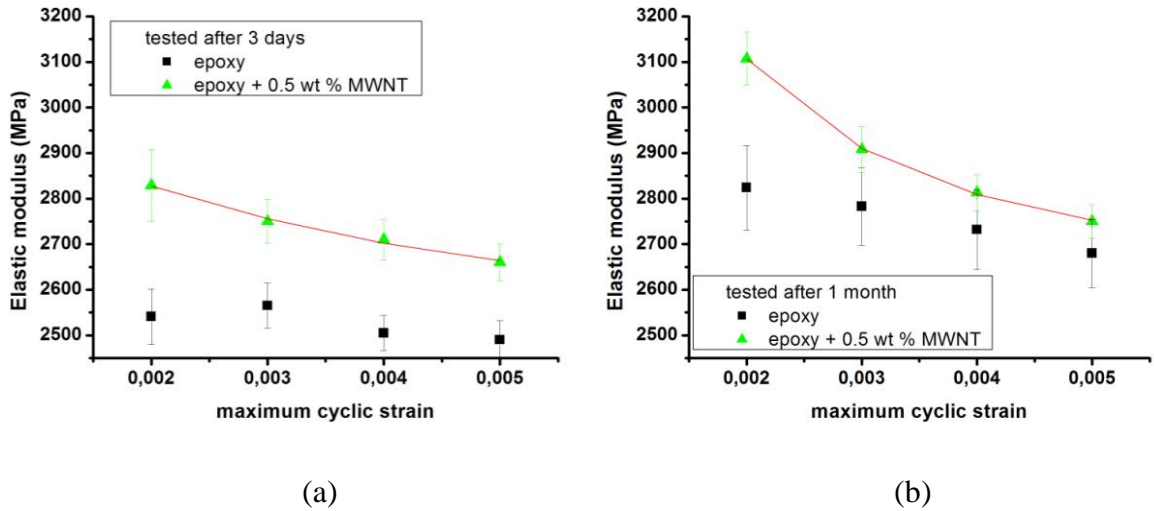

(a) (b)

Figure 1 – Elastic modulus of unfilled epoxy and composites tested after a) 3 days and b) 1 month as a function of maximum cyclic strain. Solid lines are fits to Kraus model.

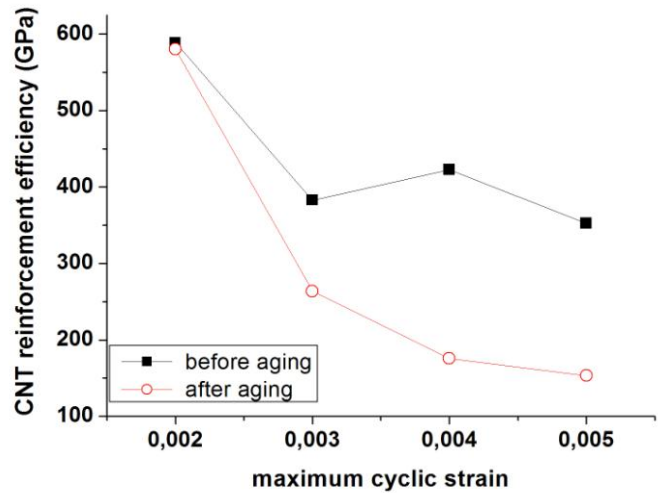

Figure 2 – CNT reinforcement efficiency as a function of maximum cyclic strain before and after aging.

|  | $E_\infty$ (MPa) | $E_0$ (MPa) | $\varepsilon_c$ |
| --- | --- | --- | --- |
| Tested after 3 days | 2,553 | 2,935 | 0.0032 |
| Tested after 1 month | 2,633 | 3,725 | 0.0017 |

Table 1 – Values of the parameters of the Kraus model fitted to the experimental tests performed on the composites after 3 days and after 1 month.